\documentclass[final]{svjour2}
\usepackage{graphicx}
\usepackage{rotating}
\usepackage{amssymb}
\usepackage{mathptmx}
\usepackage[numbers]{natbib}
\usepackage{amssymb}
\usepackage{mathrsfs}
\usepackage{amsmath}

\usepackage{graphicx}

\usepackage{dcolumn}
\usepackage{bm}
\usepackage{mathrsfs}
\usepackage{xspace} 
\makeatletter
\journalname{Journal of Low Temperature Physics}

\bibpunct{}{}{,}{s}{}{,}

\begin{document}

\newcommand{\hdblarrow}{H\makebox[0.9ex][l]{$\downdownarrows$}-}
\title{Profile likelihood ratio analysis techniques for rare event signals}

\author{J. Billard for the SuperCDMS collaboration}

\institute{Department of Physics, Massachusetts Institute of Technology, Cambridge, MA 02139, USA
\email{billard@mit.edu}}

\date{07.15.2013}

\maketitle

\begin{abstract}

The Cryogenic Dark Matter Search (CDMS) II uses crystals operated at milliKelvin temperature to search for dark matter. We present the details of the profile likelihood analysis of 140.2 kg-day exposure from the final data set of the CDMS II Si detectors that revealed three WIMP-candidate events. We found that this result favors a WIMP+background hypothesis over the known-background-only hypothesis at the 99.81\% confidence level. This paper is dedicated to the description of the profile likelihood analysis dedicated to the CDMSII-Si data and discusses such analysis techniques in the scope of rare event searches.


\end{abstract}

\section{Introduction}
\begin{figure}
\begin{center}
\includegraphics[%
  width=0.8\linewidth,
  keepaspectratio]{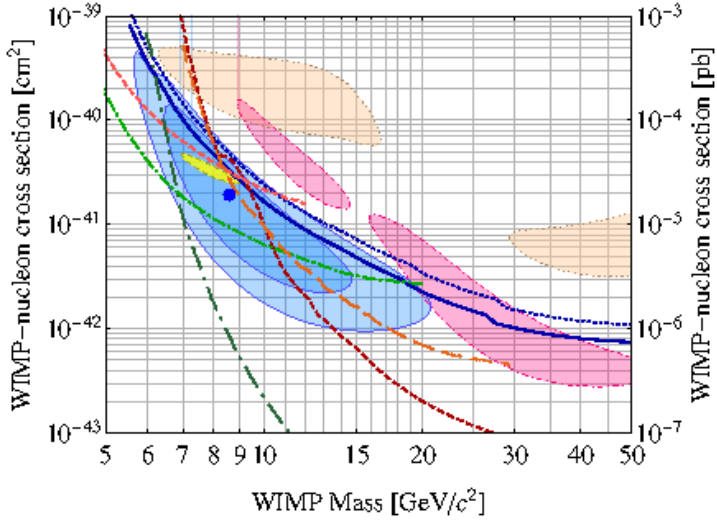}
\end{center}
\caption{(color online) Experimental upper limits (90\% confidence level) for the WIMP-nucleon spin-independent cross section as a function of WIMP mass. We show the limit obtained from the exposure analyzed in this work alone ({\it blue dotted line}), and combined with previous CDMS II Si data ({\it blue solid line}).  Also shown are limits from the CDMS II Ge standard and low-threshold analysis  ({\it dark and light dashed red}), EDELWEISS low-threshold (\emph{long-dashed orange}), XENON10 S2-only (\emph{dash-dotted green}), and XENON100  (\emph{long-dash-dotted green}). The filled regions identify possible signal regions associated with data from CoGeNT ({\it dashed yellow}, 90\% C.L.), DAMA/LIBRA  ({\it dotted tan}, 99.7\% C.L.), and CRESST  ({\it dash-dotted pink}, 95.45\% C.L.) experiments. 68\% and 90\% C.L. contours for a possible signal from these data are shown in light blue. The blue dot shows the maximum likelihood point at (8.6~GeV/c$^{2}$, $1.9\times10^{-41}$~cm$^{2}$). Figure taken from \cite{Agnese:2013rvf}, see reference therein.}
\label{fig:Contours}
\end{figure}

Direct dark matter detection consists in measuring the recoil energy from the elastic scattering of a WIMP (Weakly Interacting Massive Particle) on a detector target nucleus. The CDMS collaboration aims at measuring such nuclear recoils using cryogenic semiconductor detectors, at a temperature of about 50 mK, reading out both the phonon and the ionization signals. The discrimination between bulk nuclear recoils and background events is done by a combined use of the ionization yield measurement and timing information from the athermal phonon signals \cite{Agnese:2013rvf}.
The CDMS Collaboration recently published their result of a 140.2 kg-days exposure of Si detectors where 3 WIMP candidate events were found \cite{Agnese:2013rvf}. A dedicated profile likelihood analysis has been developed in order to further interpret these data in the context of dark matter direct detection. Following this analysis and considering the standard halo model assumptions described in \cite{Agnese:2013rvf}, we found that the results tend to be in favor of a WIMP interpretation over the background only at the 99.81\% C.L. with a most likely WIMP mass and cross section at about 8.6 GeV/c$^2$ and 1.9$\times10^{-41}$~cm$^2$ respectively, see figure~\ref{fig:Contours}\footnote{It is worth noticing that in the light of this result, the SuperCDMS collaboration has decided to consider 11 kg of Si detectors and 92 kg of Ge detectors for the proposed SuperCDMS SNOLAB experiment. That will enable the SuperCDMS experiment to probe the entire CDMS II Si allowed region presented in figure~\ref{fig:Contours} with both Si and Ge target nuclei.}. This paper describes the interest of using profile likelihood analyses when interpreting WIMP data and its application to the recently published CDMS II Si data. It is organized as follows: we first give a brief description of profile likelihood analyses, then examplify its application in the case of the CDMS II Si data and finally present relevant statistical tests based on profile likelihood statistics to further interpret the results in the scope of direct dark matter searches.

\section{Profile likelihood analysis}

Profile likelihood analyses dedicated to dark matter searches have been first introduced by the XENON collaboration \cite{Aprile:2011hx}. As opposed to a standard ``signal box'' approach, where one defines cuts such that the expected background contribution is very small, such analyses may use all the event parameter space to improve their signal acceptance. As opposed to traditional ``Optimal Interval'' or ``Optimum Gap'' methods \cite{yellin}, profile likelihood techniques are designed  to incorporate systematic uncertainties from the signal and/or background models in the derived allowed regions or upper limits in a truly frequentist framework.  \\
 Systematic uncertainties, also called nuisance parameters and denoted  $\vec{\nu}$ in the following, are profiled over in the computation of the likelihood function for each WIMP mass ($m_\chi$) and WIMP-nucleon cross section ($\sigma_{\chi-n}$). This is done by maximizing the likelihood function  over the different nuisance parameters such as:
\begin{equation}
\mathscr{L}(m_\chi,\sigma_{\chi-n},\hat{\hat{\vec{\nu}}}) = \max\limits_{\{\nu\}}\left\lbrace\mathscr{L}(m_\chi,\sigma_{\chi-n},\vec{\nu})\right\rbrace 
\end{equation}
It is therefore possible to estimate the discovery significance in taking into account relevant systematics. This is done by testing the background only hypothesis ($H_0$) on the data and trying to reject it against the signal+background hypothesis ($H_1$) using the following test statistic \cite{cowan}:
\begin{equation}
q_0 = -2\log\left\lbrace\frac{\mathscr{L}(m_\chi,\sigma_{\chi-n} = 0,\hat{\hat{\vec{\nu}}})}{\mathscr{L}(\hat{m}_\chi,\hat{\sigma}_{\chi-n},\hat{\vec{\nu}})} \right\rbrace \ \ \text{if $\hat{\sigma}_{\chi-n} > 0$}
\end{equation}
As one can deduce from such test, a large value of $q_0$ implies a large discrepancy between the two hypotheses, which is in favor of $H_1$ hence of a discovery interpretation. Following Wilk's theorem, $q_0$ asymptotically follows a $\chi^2_{2}$ distribution, see \cite{cowan} for a more detailed discussion. 

\section{Application to the CDMS II silicon data}

\begin{figure}
\begin{center}
\includegraphics[%
  width=0.45\linewidth,
  keepaspectratio]{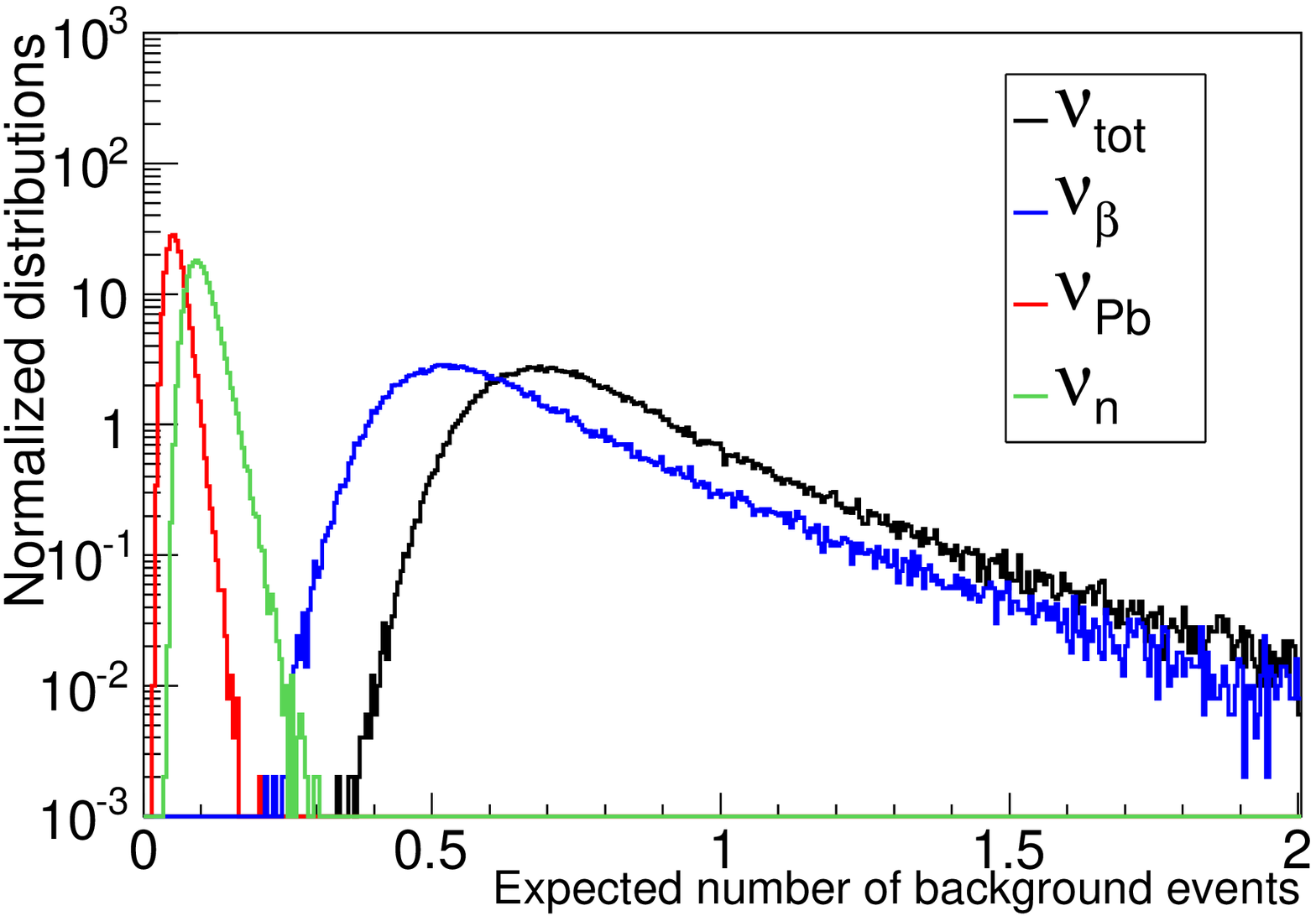}
\includegraphics[%
  width=0.45\linewidth,
  keepaspectratio]{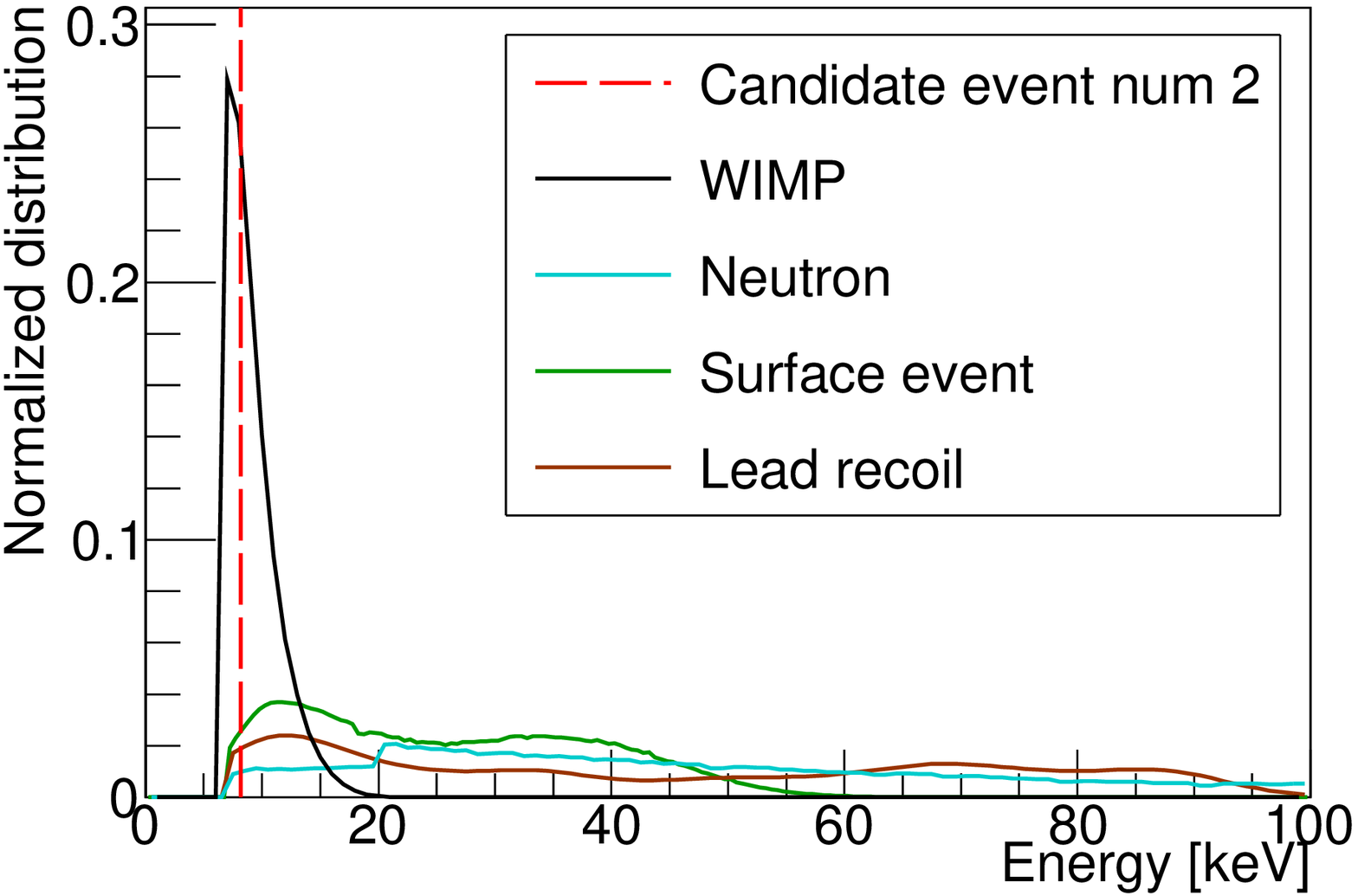}
\end{center}
\caption{(Color online) Probability Density Functions (PDF) of the expected background contaminations (left) and energy distribution of each backgrounds and the best fit WIMP model  in Detector 3 from Tower 4 (right).}
\label{fig:backgrounds}
\end{figure}

We consider data from the Si detectors using the final four runs of the full CDMS II detector installation acquired between July 2007 and September 2008. The data collected by the 8  Si detectors considered in this analysis represent a total exposure of 140.2 kg-days prior to application of the WIMP candidate selection criteria. The detectors were calibrated using $^{133}$Ba and $^{252}$Cf sources to study the detector response to gamma induced electronic recoils and nuclear recoils respectively. The candidate events were therefore required to have an ionization yield consitent with nuclear recoils and to be in the bulk region of the detectors to prevent any contamination from surface events that may suffer from reduced ionization yield. The volume fiducialisation is done using a veto outer charge electrode (radial cut) and a phonon pulse shape discrimination (top and bottom surfaces cut). The definition of the signal region discussed in  \cite{Agnese:2013rvf, kevin} led to an energy threshold of 7 keV and a nuclear recoil acceptance of about 40\% over the whole energy range of interest, {\it i.e. } [7, 100] keV. The different sources of background considered are:
\begin{itemize}
\item Surface events: The greatest source of background is the misidentification of surface electron recoils which may suffer from reduced ionization yield and hence leak into our pre-defined signal region. A Bayesian estimate of such pathological events combining calibration data and WIMP search data led us to the following  estimate:
\begin{equation}
\nu_\beta = 0.41\ ^{+0.2}_{-0.08}({\rm stat.})\ ^{+0.28}_{-0.24}({\rm syst.}) \ (68\% \ {\rm C.L.})
\end{equation}
\item Neutrons: A full Geant4 simulation of the CDMS II experiment led us to a sum of radiogenic and cosmogenic neutron contamination of:
\begin{equation}
\nu_n < 0.13 \ (90\% \ {\rm C.L.})
\end{equation}
\item $^{206}$Pb recoils from $^{210}$Po decays:  Searching for coincident $\alpha$ particles in neighboring detector we found: 
\begin{equation}
\nu_{\rm Pb} < 0.08 \ (90\% \ {\rm C.L.})
\end{equation}
\end{itemize}
The Probability Density Functions (PDF) of the  expected background contaminations for all detectors and runs combined are shown on figure~\ref{fig:backgrounds} (left panel). From these estimates of our expected background contaminations, we found that the probability  of  observing three or more events in our signal region due to background fluctuations is equal to 5.4\%.

In a profile likelihood analysis, the discrimination between background and signal events is performed using their distributions in the relevant parameter space such as Ionization yield, Phonon timing and Energy. In the case of nuclear recoils and surface events, we used multi-dimensional gaussian Kernel Density Estimators (KDE) to assess their distributions in the event parameter space. The bandwidths of the KDE were optimized using the standard maximum likelihood cross validation technique. However, due to small statistics in our event samples we checked that our analysis was much more robust against statistical fluctuations by ignoring the ionization yield and phonon timing information. Hence, the final profile likelihood analysis was performed only on candidate events in the signal region and considering only their energy distributions\footnote{Note that considering also the phonon timing and ionization yield information in the likelihood calculation could potentially improve the discrimination power between the WIMP and the background interpretation of the observed events. However, that would require higher statistics in order to get a robust and accurate estimation of the different PDFs over the whole three dimensional parameter space.}. 

During our post-unblinding checks, we found that one of our candidate events was nearly classified as a multiple scatter by the single-scatter criterion. An event is considered as a multiple scatter if any detector of the CDMS II experiment other than the one that has triggered records a phonon energy greater than 2 keV. However, following a detailed modeling of the distributions of the ``next to leading phonon energy'' of multiple and single scatters in the CDMS II experiment, we found that the probability for each event to be a multiple was respectively 96.1\%, 99.7\% and 99.7\% respectively.\\

\begin{figure}
\begin{center}
\includegraphics[%
  width=0.45\linewidth,
  keepaspectratio]{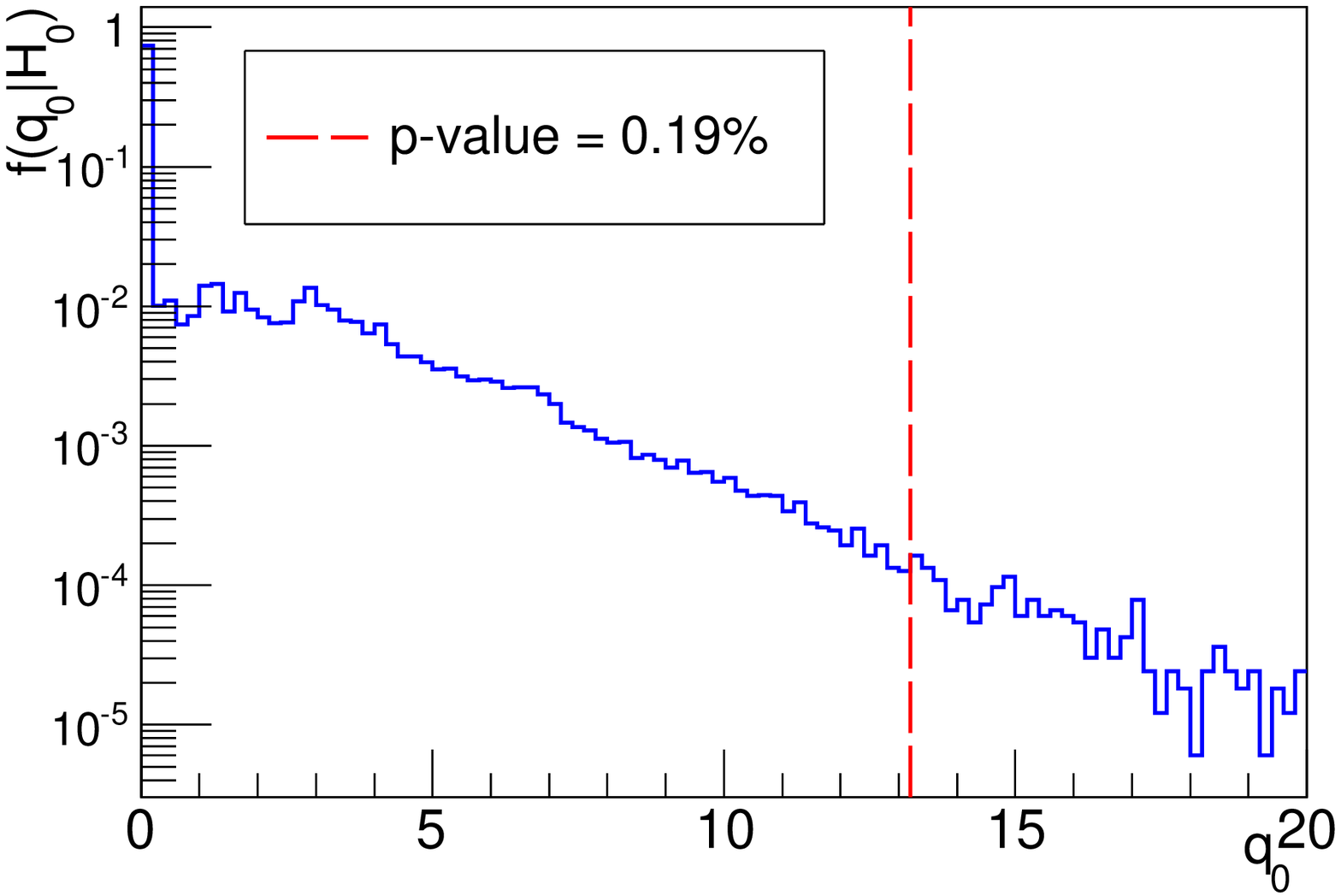}
\includegraphics[%
  width=0.45\linewidth,
  keepaspectratio]{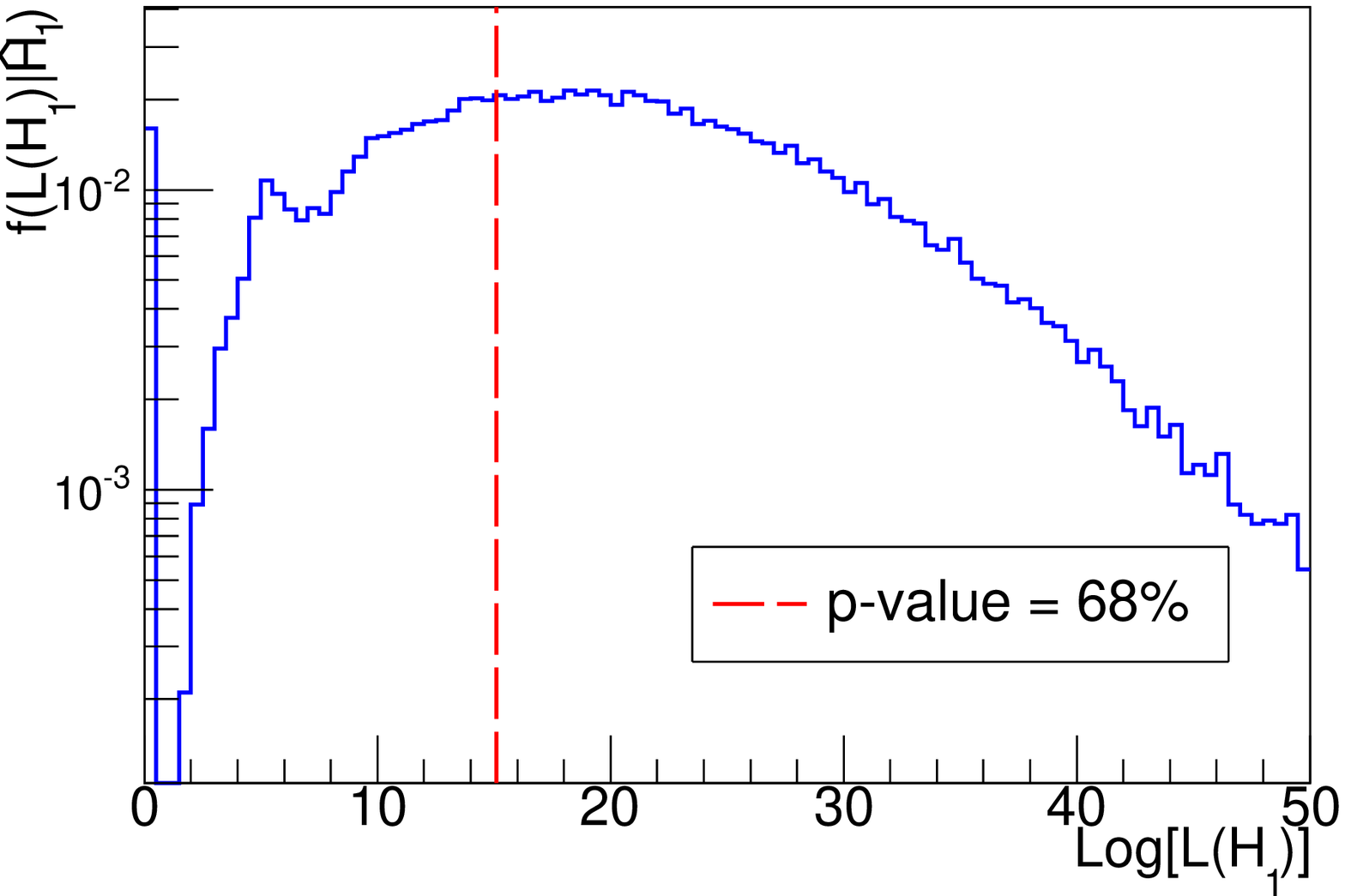}
\end{center}
\caption{(Color online) Profile likelihood ratio $q_0$ under the background only hypothesis (left) and maximum likelihood distribution under our best fit hypothesis (right).}
\label{fig:likelihoodRatio}
\end{figure}

The total likelihood function dedicated to our Si data analysis can be defined as follows,
\begin{equation}
\mathscr{L}(m_\chi,\sigma_n, \vec{\nu}) = \prod_{i=1}^{\rm N_{runs}} \prod_{j = 1}^{\rm N_{det}}\mathscr{L}_{i,j}(m_\chi,\sigma_n, \vec{\nu}_{i,j})
\end{equation}
As one can see from the above expression, the likelihood function is defined as the combination of the individual likelihood functions of each individual detectors and runs  ($\mathscr{L}_{i,j}$). $\vec{\nu}$ corresponds to the set of nuisance parameters which are the background estimates of each background population for each run and detector \footnote{With such profile likelihood analysis methods, one could also take into account astrophysical uncertainties from Dark Matter halo modeling as shown in \cite{Aprile:2011hx, Billard:2011zj}.}. \\
 Considering the background  modeling discussed in the previous section and  the event multiplicity issue, the individual likelihood functions are hence defined as, 

\begin{align}
\mathscr{L}_{i,j}(m_\chi,\sigma_n, \vec{\nu}_{i,j})  =  & \frac{\exp{\left\{-\left(\mu_{i,j} + \sum_{l = {\rm Pb}, \beta, n}\nu^{\rm SS}_{i,j,l} + \sum_{l = {\rm Pb}, \beta, n}\nu^{\rm MS}_{i,j,l} \right )\right\}} }{N_{i,j}!}  \nonumber \\
 & \times \prod_{k = 1}^{N_{i,j}}\left[ \left( \mu_{i,j}f_{s}^{i,j}(E_{k}) + \sum_{l = {\rm Pb}, \beta, n}\nu^{\rm SS}_{i,j,l}f_{l}^{i,j}(E_{k}) \right)\times f^{\rm SS}(pt_{k})\right. \nonumber \\ 
&\left. + \left(\sum_{l = {\rm Pb}, \beta, n}\nu^{\rm MS}_{i,j,l}f_{l}^{i,j}(E_{k}) \right)\times f^{\rm MS}(pt_{k}) \right] \nonumber \\
& \times \prod_{l = {\rm Pb}, \beta, n} \mathscr{L}_{i,j,l}(\nu^{\rm SS}_{i,j,l} + \nu^{\rm MS}_{i,j,l}) 
\label{eq:likelihood}
\end{align}
where the $k$ index refers to the $k$-th event observed in the $j$-th detector during the $i$-th run and the index $l$ refers to the different kind of background populations: surface events, neutrons and  $^{206}$Pb recoils while $s$ refers to the WIMP signal. The indices SS and MS refer to Single and Multiple Scatter respectively. $pt_k$ and $E_k$ correspond respectively to the next to leading phonon energy ($pt$) from non-triggered detectors and the recoil energy of the $k$-th event. As one can see from  equation \ref{eq:likelihood}, the individual likelihood  have three main components which are: a Poisson term to take into account statistical fluctuations, a term with  the event  distributions to discriminate between signal and backgrounds, and the likelihood functions for each nuisance parameters that are given by the background estimates PDF for each run and detectors. Such a likelihood function is then characterized by 2 parameters of interest ($m_\chi,\sigma_{\chi-n}$) and a total of 192 nuisance parameters, {\it i.e.} 6 for each individual likelihood functions. \\

\section{Test statistics and $p$-values}

In order to further assess the validity of the Dark Matter interpretation of the data, it is of first importance to estimate the statistical relevance of the observed signal. To do so, two statistical tests are required to estimate both the signal significance and the goodness of fit of the favored WIMP hypothesis.\\
The statistical significance of the WIMP + background hypothesis is done using the standard profile likelihood ratio test statistic discussed above.  Because of very low statistics, we  computed the distribution $f(q_0|H_0)$ of $q_0$ under $H_0$ from 170,000 Monte Carlo simulations of our known background model taking into account systematic uncertainties in the background  estimates from the PDFs shown in figure~\ref{fig:backgrounds} (left). The resulting distribution is shown on figure~\ref{fig:likelihoodRatio} as the blue histogram while the observed value from the data $q_0^{\rm obs} = 13.2$ is shown as the red dashed line. We found a $p$-value of $p_0 = 0.19\%$ implying that the WIMP+background hypothesis is being favored over the background only hypothesis at 99.81\%. \\

Likelihood ratio tests can only tell us which of the two hypotheses is being favored over the other, but do not give us any information on whether the favored hypothesis seems to correspond to the observed data. Therefore, it is compulsory to  test the goodness of fit of the favored WIMP+background hypothesis. This has been done by computing the maximum likelihood distribution from 60,000 Monte Carlo simulations of our known background model combined with a 8.6 GeV/c$^2$ and a 1.9$\times$10$^{-41}$ cm$^2$ WIMP model contribution. The resulting distribution   as well as the observed value are presented in figure~\ref{fig:likelihoodRatio} (right). The $p$-value of this goodness of fit test is found to be equal to 68\% suggesting that our three observed events are very well fitted by our best-fit WIMP+background model.

\section{Conclusions}

We have shown that  profile likelihood analyses are well-suited to interpret observed data in the scope of any rare event searches experiments. As an illustration, we considered the case of the recently published CDMS II Si data that has observed three candidate events. A careful modelling of the backgrounds and their associated systematics combined with a dedicated profile likelihood analysis have enable us to further interpret the Silicon results. The main drawback of such analyses is that they depend on the accuracy of the estimation of the backgrounds. This is particularly important to keep in mind for rare event searches where background estimates are inferred from very low statistics.

\begin{acknowledgements}
The CDMS collaboration gratefully acknowledges the contributions of numerous 
engineers and technicians; we would like to especially thank Dennis Seitz, 
Jim Beaty, Bruce Hines, Larry Novak, 
Richard Schmitt and Astrid Tomada.  In addition, we gratefully acknowledge assistance 
from the staff of the Soudan Underground Laboratory and the Minnesota Department of Natural Resources. 
This work is supported in part by the 
National Science Foundation the Department of Energy,  the Swiss National 
Foundation, the NSERC Canada and by MULTIDARK.
\end{acknowledgements}


\begin{thebibliography}{99}

\bibitem{Agnese:2013rvf} 
  R.~Agnese {\it et al.}  [CDMS Collaboration],
  [arXiv:1304.4279 [hep-ex]].

\bibitem{Aprile:2011hx} 
  E.~Aprile {\it et al.}  [XENON100 Collaboration],
  Phys.\ Rev.\ D {\bf 84}, 052003 (2011)
  [arXiv:1103.0303 [hep-ex]].

\bibitem{yellin}
S. Yellin, Phys. Rev. D {\bf 66}, 032005 (2002) 

\bibitem{cowan}
G. Cowan, K. Cranmer, E. Gross, O. Vitells, Eur. Phys. J. {\bf C71}, 1554 (2011)

\bibitem{kevin}
K. A. McCarthy, Ph.D. thesis, Massachusetts Institute of Technology (2013).

\bibitem{Billard:2011zj} 
  J.~Billard, F.~Mayet and D.~Santos,
  Phys.\ Rev.\ D {\bf 85}, 035006 (2012)

\end{thebibliography}
\end{document}